\documentclass[showpacs,twocolumn,floatfix]{revtex4}
\usepackage{graphicx}
\usepackage{dcolumn}
\usepackage{bm}
\usepackage{amssymb}
\usepackage{amsmath}
\usepackage{epsf}
\usepackage{subfigure}
\usepackage{epstopdf}%
\usepackage{amsfonts}
\usepackage{color}

\begin{document}

\title{Pulse propagation in tapered granular chains: An analytic study}
\author{Upendra Harbola$^1$, Alexandre Rosas$^2$, Massimiliano Esposito$^{1,3}$ and Katja Lindenberg$^1$}
\affiliation{${}^1$Department of Chemistry and Biochemistry, University of California San Diego, La
Jolla, California 92093-0340, USA.\\
${}^2$Departamento de F\'{\i}sica, CCEN, Universidade Federal da Para\'{\i}ba, Caixa Postal
5008, 58059-900, Jo\~ao Pessoa, Brazil.\\
${}^{3}$Center for Nonlinear Phenomena and Complex Systems,
Universit\'e Libre de Bruxelles, Code Postal 231, Campus Plaine, B-1050 Brussels, Belgium.\\}

\date{\today}

\begin{abstract}
We study pulse propagation in one-dimensional tapered chains of spherical
granules.  Analytic results for the pulse velocity and other pulse features are
obtained using a binary collision approximation.  Comparisons with numerical results show that the
binary collision approximation provides quantitatively accurate analytic results for these chains.
\end{abstract}


\maketitle 

\section{Introduction}
\label{introduction}

The study of pulse propagation through granular media has attracted a great deal of attention
for several reasons. Apart from addressing fundamental problems of pulse
dynamics in the presence of highly nonlinear interactions, it also has direct
practical application. Pulse propagation has mostly been studied in monodisperse
one-dimensional chains of spherical granules.
Nesterenko first showed~\cite{nesterenko,nesterenko-1} that an initial impulse
at an edge of a granular chain in the absence of precompression can result in
solitary waves propagating through the medium. Since then, this system
and variants thereof have been a testbed of extensive theoretical and experimental
studies~\cite{nesterenko-book,application,alexandremono,alexandre,jean,wangPRE07,rosasPRL07,rosasPRE08,sen-review,costePRE97,nesterenkoJP94,sokolowEPL07,robertPRL06,wu02,robertPRE05,sokolowAPL05,senPHYA01,meloPRE06,jobGM07}.

The chain model has recently been
generalized to study the (sometimes profound) effects on pulse propagation
of dissipation~\cite{application,alexandre,rosasPRL07,rosasPRE08} and of
polydispersity in the structure and mass of the
granules~\cite{wangPRE07,sen-review,robertPRL06,wu02,robertPRE05,sokolowAPL05,senPHYA01}.
In particular, polydispersity is frequently introduced in a regular fashion such as in tapered chains
(TCs), in which the size and/or mass of successive granules systematically decreases or increases.
Polydispersity is also introduced by distributing masses randomly, by ``decorating" chains with
small masses regularly or randomly placed among larger masses, and by optimizing grain distribution
for particular purposes.  For example, considerable recent activity has focused on pulse propagation in
``forward tapered chains"~\cite{robertPRL06,wu02,robertPRE05,sokolowAPL05,senPHYA01}, with some
results corroborated experimentally~\cite{meloPRE06,jobGM07}.  In these chains each
granule is smaller than the preceding granule according to a systematic pattern.  Among the main
results of these studies is that the energy imparted to the first granule is redistributed among
increasingly larger numbers of granules as the pulse propagates, thus attenuating the impulse on 
subsequent granules as time proceeds.  This is a desirable behavior in the context of shock absorption. 
In any case, almost all known results are numerical and therefore predictively
limited. While a few specific realizations have been experimentally verified, we have found 
few analytic results for pulse propagation in polydisperse systems~\cite{robertPRL06}.
Analytic work is in principle possible
(albeit in practice often extremely difficult) if a continuum approximation is valid.  This approach was
first implemented successfully for monodisperse granular chains, where the resulting continuum
equations can be (approximately) solved
analytically~\cite{nesterenko,nesterenko-1,nesterenko-book,alexandre,jean}.
Even when a continuum
formulation is appropriate, in many cases the resulting equations can not be
solved~\cite{rosasPRL07,rosasPRE08}. 

In this work we develop a converse approach which focuses on the fact that the propagating
front in granular chains is often narrow.
Our theory invokes a \emph{binary collision} approximation which leads to analytic
expressions for various quantities that characterize the propagating pulse.  In an earlier
paper we introduced a binary collision approximation for the calculation of the pulse velocity 
in monodisperse chains~\cite{alexandremono}.  Interestingly, although it is ``opposite" to the continuum
approximation, we found that the binary collision model produces results that also capture the dynamics
of pulse propagation in such chains.  Here we generalize this approach to forward and
backward TCs, that is, to chains in which successive granules increase in size or decrease in size.
The high quality of the approximation is ascertained by comparison with
numerical integration of the full equations of motion for two
tapering protocols.  In one, the granular radius increases or decreases linearly (backward or forward
``linearly" TC), while in the second the radius increases or decreases exponentially (backward
or forward ``exponentially" TC).

In Sec.~\ref{model} we introduce the granular chain model and lay out the binary approximation
procedure and its caveats.  The analytic results for the backward TCs and comparisons for these
chains with numerical integration results are presented in Sec.~\ref{backward}.  The results and
comparisons for the forward TCs are presented in Sec.~\ref{forward}. Section~\ref{summary}
provides a summarizing closure.

\section{The model and the binary approximation}
\label{model}

We consider chains of granules all made of the same material of density $\rho$.  
When neighboring granules collide, they repel each other according to
the power law potential
\begin{equation}
 \label{hertz-1}
 V=\frac{a}{n} r_k^\prime |y_k-y_{k+1}|^n.
\end{equation}
Here $y_k$ is the displacement of granule $k$ from its position at the beginning of the collision,
and $a$ is a constant determined by Young's modulus and Poisson's ratio~\cite{landau,hertz}.
The exponent $n$ is $5/2$
for spheres (Hertz potential), which we use in this paper~\cite{hertz}. We have defined 
\begin{equation}
  r_k^\prime = \left(\frac{2 R_k^\prime R_{k+1}^\prime}{R_k^\prime +R_{k+1}^\prime }\right)^{1/2},
\end{equation}
where $R_k^\prime$ is the principal
radius of curvature of the surface of granule $k$ at the point of contact.
When the granules do not overlap, the potential is zero.
The equation of motion for the $k$th granule is
\begin{eqnarray}
 \label{EOM-1}
 M_k\frac{d^2y_k}{d{\tau}^2}&=& {a}{r}_{k-1}^\prime(y_{k-1}-y_{k})^{n-1}\theta(y_{k-1}-y_k)\nonumber\\
&-& {a}{r}_{k}^\prime(y_k-y_{k+1})^{n-1}\theta(y_k-y_{k+1}),
\end{eqnarray}
where $M_k=(4/3)\pi \rho (R_k^\prime)^3$.  The Heaviside function
$\theta(y)$ ensures that the elastic interaction between grains only exists if they are in contact.  
Initially the granules are placed along a line so that they
just touch their neighbors in their equilibrium positions (no
precompression), and
all but the leftmost particle are at rest. The initial velocity of
the leftmost particle ($k=1$) is $V_1$ (the impulse). 
We define the dimensionless quantity
\begin{equation}
\alpha\equiv \left[ \frac{M_1V_1^2}{a \left(R_1^\prime\right)^{n+1/2}}\right]
\end{equation}
and the rescaled quantities $x_k$, $t$, $m_k$, and $R_k$ via the relations
\begin{eqnarray}
y_k = R_1^\prime\alpha^{1/n} x_k, &\qquad&
\tau = \frac{R_1^\prime}{V_1} \alpha^{1/n} t, \nonumber\\
R_k^\prime = R_1^\prime R_k, &\qquad& M_k=M_1 m_k.
\end{eqnarray}
Equation~(\ref{EOM-1}) can then be rewritten as
\begin{eqnarray}
  m_k \ddot{x}_k &=&
 r_{k-1}(x_{k-1} - x_k)^{n-1} \theta (x_{k-1} -
x_k) \nonumber\\
&&-  r_k (x_k - x_{k+1})^{n-1} \theta (x_k -
x_{k+1}),
\label{eq:motion_rescaled}
\end{eqnarray}
where a dot denotes a derivative with respect to $t$, and
\begin{equation}
 \label{r-prime}
r_k= \left(\frac{2 R_kR_{k+1}}{R_k+R_{k+1}}\right)^{1/2}.
\end{equation}
The rescaled initial velocity is unity, i.e., $v_1(t=0)=1$. 
The velocity of the $k$-th granule in unscaled variables is simply $V_1$ times its velocity in the
scaled variables.

The equations of motion can be integrated numerically, and we do so for our TCs.
In each chain we observe that the initial impulse quickly settles into a propagating pulse
whose front typically involves only three granules.  In the
backward TCs the narrow pulse is ``clean" in the sense that granules behind it are
reflected back and thus do not participate in forward energy propagation.  In the forward
TCs the narrow front carries a tail of moving granules behind
it, but this tail does not affect the propagation dynamics of the narrow front.
From the numerical integration results we are able to extract quantities such as the pulse peak
location, amplitude, and velocity as functions of granule index and of time.  These are
the quantities to be compared with those obtained from the binary collision theory. 

The binary collision approximation is based on the assumption that the transfer of energy along the
chain occurs via a succession of two-particle collisions. First, particle $k=1$
of unit velocity
collides with initially stationary particle $k=2$, which then acquires a velocity $v_2$
and collides with stationary
particle $k=3$, and so on.  The velocities after each collision can easily be obtained from
conservation of energy and momentum.  The velocity $v_{k+1}$ of granule $k+1$ after the collision
with granule $k$ is always positive, so this particle goes on to
collide with the next one. On the other hand, the velocity of granule $k$ after the collision
may be positive or negative depending on the
direction of tapering of the chain (it would be zero for a monodisperse chain).
From these velocities, a number of other results follow which
we outline here, but implement for particular chains in later sections. 

After the collision of granules $k$ and $k+1$, the latter emerges with velocity
\begin{equation}
 \label{velo-k+111}
 v_{k+1}=\frac{2v_k}{1+\frac{\displaystyle m_{k+1}}{\displaystyle m_k}}.
\end{equation}
This result can be implemented recursively to obtain
\begin{equation}
  v_k = \prod_{k^\prime=1}^{k-1} \frac{2}{1+\frac{\displaystyle m_{k^\prime+1}}
{\displaystyle m_{k^\prime}}}.
\label{recv}
\end{equation}
To actually evaluate this product we need to implement a tapering protocol, cf. below.  

We wish to use these results to calculate the time it takes the pulse to move along the
chain and, more specifically, the time it takes the pulse to arrive at the $k$th granule. 
For this purpose, it is convenient to introduce the difference variable 
\begin{equation}
  z_{k}=x_k-x_{k+1}.
\end{equation}
The equation of motion for the difference variable is obtained by subtracting
the equations of motion of the two granules during a collision, cf. Eq.~(\ref{eq:motion_rescaled}),
\begin{eqnarray}
 \label{eom-11}
\ddot{x}_k &=& -\frac{r_k}{m_k}(x_k-x_{k+1})^{n-1}\nonumber\\
\ddot{x}_{k+1} &=& \frac{r_k}{m_{k+1}}(x_k-x_{k+1})^{n-1},
\end{eqnarray}
which directly leads to
\begin{equation}
\label{eq-z1}
\mu_k\ddot{z}_{k}=-{r}_k z_{k}^{n-1}.
\end{equation}
Here $\mu_k=m_k m_{k+1}/(m_k+ m_{k+1})$ is the reduced mass. Equation~(\ref{eq-z1}) is the
equation of motion of a particle of mass $\mu_k$ in the potential $(r_k/n)z_k^n$ defined for $z_k\geq 0$.
The initial condition $\dot{z}_k(0)$ is simply $v_k$ since the velocity of granule $k+1$ is zero
before the collision.  Therefore, from Eq.~(\ref{recv}) we can write
\begin{equation}
\label{initial-condition-binary}
\dot{z}_{k}(0)=
\prod_{k^\prime=1}^{k-1}
\frac{2}{\left(1+
\frac{\displaystyle m_{k^\prime+1}}{\displaystyle m_{k^\prime}}\right)
} .
\end{equation}
The energy conservation condition
\begin{equation}
  \frac{1}{2} \dot{z}_k^2(t) +\frac{r_k}{n\mu_k} z_k^n(t) = \frac{1}{2} \dot{z}_k^2(0)
\label{max1}
\end{equation}
leads to
\begin{equation}
  \dot{z}_k(t) = \left( \dot{z}_k^2(0) -\frac{2r_k}{n\mu_k}z_k^n(t) \right) ^{1/2}.
\label{max}
\end{equation}

We say that the pulse arrives at granule $k$ when the velocity of granule $k$ surpasses that of
granule $k-1$, and that it moves on to granule $k+1$ when the velocity of the $(k+1)$st granule
surpasses that of the $k$th granule.  The residence time $T_k$ on granule $k$ is
the time that granule $k$ takes to transfer the pulse from $k-1$ to $k+1$, and is given by
\begin{equation}
T_k = \int_0^{z_k^{max}} \frac{dz_k}{\dot{z_k}}
 = \int_0^{z_k^{max}}\frac{dz_{k}}{\left(\dot{z}_{k}^2(0)-
\frac{\displaystyle 2r_k}{\displaystyle n\mu_k}z_{k}^n\right)^{1/2}},
\end{equation}
where $z_k^{max}$ is the compression when the velocities of particles $k$ and $k+1$ are equal
[which is also the maximum compression, cf. Eq.~(\ref{max1})],
\begin{equation}
  z_k^{max} = \left( \frac{n \mu_k}{2r_k} \dot{z}_k^2(0)\right) ^{1/n}.
\end{equation}
The integral can be performed exactly to yield
\begin{equation}
 \label{compress-time}
 T_k = \sqrt{\pi}\left(\frac{n\mu_k}{2r_k}\right)^{1/n} [\dot{z}_{k}(0)]^{(-1+2/n)}
\frac{\Gamma(1+1/n)}{\Gamma(1/n+1/2)}.
\end{equation}
Finally, the total time taken by the pulse to pass the $k$th granule is obtained by summing $T_k$,
\begin{equation}
t=\sum_{k^\prime=1}^k T_{k\prime}.
\label{totaltime}
\end{equation}
For a monodisperse chain $\mu_k = 1/2$, $r_k = 1$ and $\dot{z}_k(0)= 1$, and
Eq.~(\ref{compress-time}) reduces to the result obtained in~\cite{alexandremono}.
To evaluate these times explicitly for tapered chains
we need to specify a tapering protocol.

\section{Backward Tapered Chains}
\label{backward}

We proceed to implement these results for various tapered chains of spherical granules ($n=5/2$), 
starting with backward
tapering in this section.  Two forms of tapering are considered.

\subsection{Linearly tapered chains}
\label{linear1}
In backward linearly TCs the radii of the granules grow linearly as
${R_k= 1+S(k-1)}$, where $S>0$ is a fixed parameter.  The ratios of
the radii and of the masses of two successive granules then are 
\begin{eqnarray}
\label{ratio-linear}
\frac{R_k}{R_{k+1}}&=& 1-\frac{S}{1+Sk}\nonumber\\
\frac{m_k}{m_{k+1}}&=& \left(1-\frac{S}{1+Sk}\right)^3.
\end{eqnarray}
The ratio of the masses can be substituted into Eq.~(\ref{recv}),
\begin{equation}
v_k = \prod_{k^\prime=1}^{k-1} \frac{2}{1+\left( 1-\frac{\displaystyle S}
{\displaystyle 1+Sk^\prime}\right)^{-3}}.
\label{fullvk}
\end{equation}
A large-$k$ limit can be evaluated by calculating the derivative of $\ln v_k$ with respect to $k$,
retaining leading terms in a Taylor expansion when both $k$ and $Sk \gg 1$, and exponentiating. 
This leads to the asymptotic scaling relation
\begin{equation}
 \label{vel-ampl-lin-large}
v_k \sim k^{-3/2}.
\end{equation}
The amplitude of the velocity pulse decreases with increasing grain number,
as would be expected since the granules grow in size and mass.  

The associated parameters for a backward linearly TC are given by
\begin{eqnarray}
\mu_k &=&  (1+Sk)^3 \left(\frac{\left(1-\frac{\displaystyle S}{\displaystyle 1+Sk}\right)^3}
{1+\left(1-\frac{\displaystyle S}{\displaystyle 1+Sk}\right)^3}\right) 
\label{mk-ak-linear-1}\\ \nonumber\\
r_k &=& (1+Sk)^{1/2} \left(\frac{1-\frac{\displaystyle S}{\displaystyle
1+Sk}}{1-\frac{\displaystyle S/2}{\displaystyle 1+Sk}}\right)^{1/2}. \label{mk-ak-linear-2}
\end{eqnarray}
Using Eqs.~(\ref{vel-ampl-lin-large}),
(\ref{mk-ak-linear-1}) and (\ref{mk-ak-linear-2})
in Eq.~(\ref{compress-time}),
we find that in the large $k$ limit $T_k$ varies with $k$ according to
\begin{equation}
\label{rel-t-k}
T_k\sim k^{13/10}.
\end{equation}
The total time $t$ taken by the pulse to reach the $k$th granule is obtained from
Eq.~(\ref{totaltime}).  For large $k$, $t$ varies as
\begin{equation}
\label{time-with-k-binary}
t\sim k^{23/10}.
\end{equation}
This growth with $k$ again reflects the slowing down of the pulse as the granules become
more massive.  Using Eq.~(\ref{time-with-k-binary})
in Eq.~(\ref{vel-ampl-lin-large}), the decay in time of the pulse amplitude is found to be
\begin{equation}
\label{ampl-timelinear}
v(t)\sim t^{-15/23}.
\end{equation}

Finally, in order to compute the speed of the pulse in space,
we note that at $t=0$ the position of each granule in the uncompressed chain is
\begin{equation}
X_0(k) = Sk^2 +2k(1-S) +S -2,
\end{equation}
and when the pulse has moved to the 
$k$-th granule its position is given by 
\begin{equation}
\label{position}
X(k,t)=X_0(k)+x_k(t).
\end{equation}
For large $k$ values, the static contribution $X_0(k)$ dominates and the position of the pulse is
\begin{equation}
X(t)\sim k^2\sim t^{20/23}.
\label{positiont}
\end{equation}
The speed $c(t)$ of the pulse therefore varies in time as 
\begin{equation}
 \label{speed}
c(t)=\frac{dX}{dt}\sim t^{-3/20}. 
\end{equation}

Thus, the binary collision approximation in the backward linearly TC leads to a power
law dependence in both $k$, Eq.~(\ref{time-with-k-binary}),
and $t$, Eq.~(\ref{ampl-timelinear}), for the pulse amplitude,
for the position (\ref{positiont}), and for the speed~(\ref{speed}).
Note that these asymptotic decays are independent of the tapering parameter $S$.

\begin{figure}[h]
\centering
\rotatebox{0}{\scalebox{.30}{\includegraphics{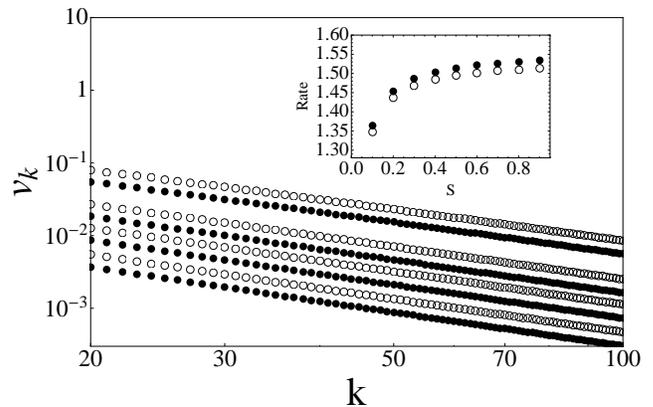}}}
\caption{The numerical integration data (solid circles) is compared with the binary collision approximation
(open circles) for the decay of the velocity pulse amplitude with $k$
for backward linearly TCs with
with $S=0.2$, $0.4$, $0.6$, and $0.9$, from
top to bottom. The inset shows the decay exponent, whose theoretical
value approaches $3/2$.  The numerical integration results also approach an $S$-independent value
that is slightly larger than but close to the theoretical value.}
\label{figure1}
\end{figure}

The comparison of the predictions of the binary collision approximation with the numerical integration
of Eq.~(\ref{eq:motion_rescaled}) is carried out as follows. 
First, consider the decay of the amplitude of the
velocity pulse as a function of $k$.  In Fig.~\ref{figure1} we show $v_k$ vs $k$ from numerical
integration of the equations of motion 
(solid circles) and as calculated from the binary collision theory, Eq.~(\ref{fullvk})
(open circles).  The comparison is made for various values of $S$, as indicated in the caption. Note
that in some regimes of this figure the decay of the
pulse amplitude \emph{is} $S$-dependent, as found in Eq.~(\ref{fullvk}) and also
from the numerical integration;
the $S$-independent power law decay~(\ref{vel-ampl-lin-large}) is only
valid when $Sk\gg 1$.  Figure~\ref{figure1}
reveals the single weakest aspect of the theory in that it shows substantial differences in the
absolute magnitude of the pulse amplitude between the theoretical and numerical
integration results, but the rate
of decay of $v_k$ is captured very well, within $\approx1\%-2\%$, as emphasized by the inset.  The
inset also shows the approach to the $S$-independent power $3/2$ for large $k$, as predicted in
Eq.~(\ref{vel-ampl-lin-large}).

\begin{figure}[h]
\centering
\rotatebox{0}{\scalebox{.30}{\includegraphics{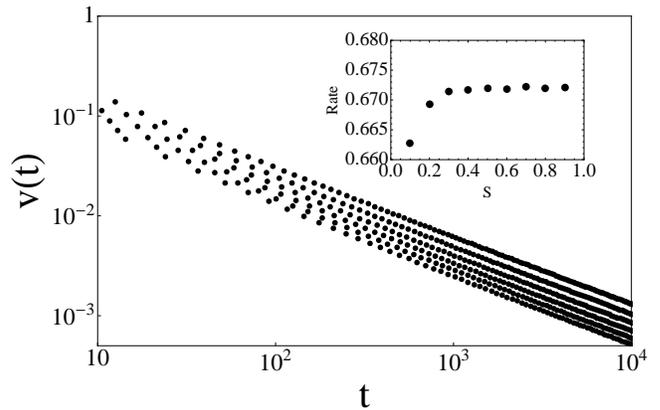}}}
\caption{The decay of the velocity pulse amplitude with time for backward linearly TCs with
$S=0.4$ to  $0.9$ in steps of $0.1$ from top to
bottom, as obtained from numerical integration. The inset shows the  rate of decay for different $S$ values.
The long-time binary prediction is $15/23=0.652\ldots$, within $2-3\%$ of the numerical integration
results.}
\label{figure2}
\end{figure}

\begin{figure}[h]
\centering
\rotatebox{0}{\scalebox{.30}{\includegraphics{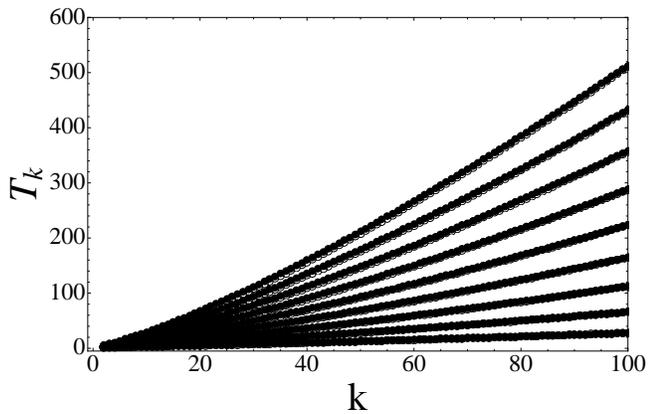}}}
\caption{Residence time of the pulse on granule $k$ for backward linearly TCs.
Comparison between the binary collision
approximation (open circles, barely visible because of the agreement between theory and numerical
integration) and direct numerical integration of the equations of motion (solid
circles). From bottom to top, $S=0.1$ to $0.9$ in steps of $0.1$.}
\label{figure2a}
\end{figure}

\begin{figure}[h]
\centering
\rotatebox{0}{\scalebox{.30}{\includegraphics{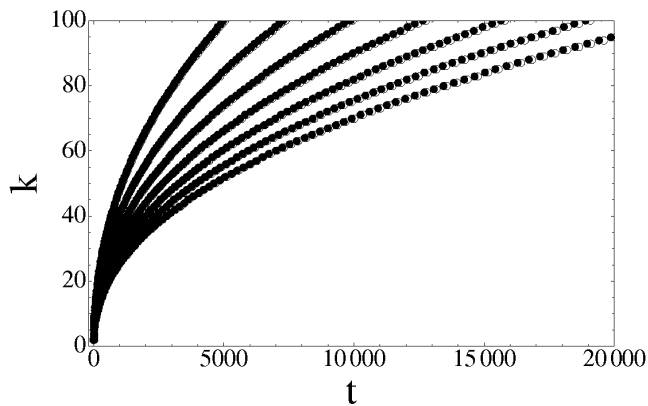}}}
\caption{Position of the pulse in units of grains versus time for backward linearly TCs.
Comparison between the binary collision
approximation (open circles, barely visible because of the agreement between theory and numerical
integration) and direct numerical integration of the equations of motion (solid
circles). From top to bottom, $S=0.3$ to $0.9$ in steps of $0.1$.}
\label{figure3}
\end{figure}

The decay of the velocity pulse in time as obtained from numerical integration
is shown in Fig.~\ref{figure2}. The power law decay~(\ref{ampl-timelinear}) predicted from the binary theory
is in excellent agreement with the numerical data. In the inset we show the power law
exponent for different $S$ values. The long-time
value of the exponent $15/23=0.652\ldots$ obtained from the theory is in very good agreement
(within $2\%-3\%$) with the observed data.
Furthermore, gratifying agreement is found if we compare the time $T_k$ obtained from
Eq.~(\ref{compress-time}) with numerical integration results, as seen in Fig.~\ref{figure2a}.
The time $t$ that the pulse takes to pass by granule $k$ is also very well reproduced by the theory,
cf. Fig.~\ref{figure3}.

In Table \ref{tab:kexp1} the agreement between the theory and the integration results is made explicit.
We list the results of fitting numerical integration
data for much longer chains ($N>400$)
ith a power law form $k=at^b$. The numerical uncertainties (standard deviations) 
are in the third decimal place.
For small values of $S$ the data does not reach the values of $k$ that are needed for the asymptotic
behavior $t^{10/23}$ of Eq.~(\ref{time-with-k-binary}) to be valid ($Sk \gg 1$). 
For the larger values of $S$ the asymptotic
behavior is approached, and here the agreement with the predicted
asymptotic value $b=10/23=0.435\ldots$ becomes evident.
Since Eq.~(\ref{time-with-k-binary}) is valid only in the large $k$ regime, 
in these fits we have only used the information for granules $300-400$.

\begin{table}
  \centering
\begin{tabular}{|l|l|l|}
  \hline
  S & theory & numerics\\ \hline
  0.3 &  0.439 &  0.438\\ \hline
  0.4 &  0.438 &  0.437\\ \hline
  0.5 &  0.437 &  0.436\\ \hline
  0.6 &  0.436 &  0.436\\ \hline
  0.7 &  0.436 &  0.436\\ \hline
  0.8 &  0.436 &  0.435\\ \hline
  0.9 &  0.436 &  0.435\\ \hline
\end{tabular}
  \caption{Backward linearly TCs: Values
of $b$ obtained from the binary collision approximation and from the numerical
integration of the equations of motion in the asymptotic expression $k\sim t^b$ for
the position of the pulse (in units of grains) as a function of
time. The theoretical asymptotic value of $b$ is $10/23=0.435\ldots$.}
  \label{tab:kexp1}
\end{table}

\subsection{Exponentially tapered chains}
\label{exponential1}

A backward exponentially TC is one in which the ratio of radii of successive granules is
constant,
\begin{equation}
 \label{constant-ratio}
\frac{R_k}{R_{k+1}} = 1-q,
\end{equation}
where $0 < q < 1$. The radii of granules thus increase in geometric progression. The radius of the
$k$th granule is $R_k=(1-q)^{1-k}$, and the mass ratio of two consecutive granules is
\begin{equation}
 \label{mass-ratio-1}
\frac{m_k}{m_{k+1}} =(1-q)^3.
\end{equation}
This is the case that has been considered 
in many previous studies~\cite{robertPRL06,wu02,robertPRE05,sokolowAPL05,senPHYA01}.
This mass ratio in Eq.~(\ref{recv}) leads to the velocity amplitude
\begin{equation}
\label{eq:vgeometric}
  v_k = A(q) e^{-k \ln A(q)},
\end{equation}
where
\begin{equation}
  A(q) = \frac{1}{2} \left[ 1 + \left( 1 - q \right)^{-3} \right],
  \label{eq:Adef}
\end{equation}
and to the associated parameters 
\begin{eqnarray}
\label{mu-binary}
 \mu_k &=& \frac{(1-q)^{-3(k-1)}}{[1+(1-q)^3]} \\ \nonumber\\
r_k &=& (1-q/2)^{-1/2} (1-q)^{-(k-1)/2}.
\label{ak-binary}
\end{eqnarray}
Substituting  Eqs.~(\ref{eq:vgeometric}), (\ref{mu-binary}), and (\ref{ak-binary}) in
Eq.~(\ref{compress-time})
and following the steps implemented for the linearly TC we find that the time $t$ for the pulse to
reach the $k$th granule now grows exponentially with $k$ for large $k$,
\begin{equation}
 \label{time-k-binary}
t \sim e^{\eta(q)k},
\end{equation}
 where 
\begin{equation}
 \label{Bq}
\eta(q) = \ln\left(\frac{[A(q)]^{1/5}}{1-q}\right).
\end{equation}

Combining Eq.~(\ref{eq:vgeometric}) with Eq.~(\ref{time-k-binary}) leads
to the decay of the pulse amplitude as a function of time,
\begin{equation}
 \label{vel-t-binary}
v(t)\sim t^{-f(q)},
\end{equation}
with
\begin{equation}
 \label{bq-geometric}
f(q)= \frac{1}{\eta(q)}\ln A(q).
\end{equation}
The position of the $k$th granule at time $t$ is 
\begin{equation}
 \label{pos-kth}
X(k,t) = \frac{2-q}{q}\left((1-q)^{-k+1}-1\right)+ x_k(t).
\end{equation}
For large $k$, we can again ignore the contribution due to the displacement compared to the
dominant static part. This allows us to approximate the pulse position as
\begin{equation}
\label{pulse-position-binary} 
X(k)\sim \frac{2-q}{q(1-q)}e^{-k \ln (1-q)},
\end{equation}
which implies that
\begin{equation}
 \label{position-pulse-binary}
X(t)\sim t^{1-f(q)/5}
\end{equation}
and hence the pulse velocity in time decays as
\begin{equation}
 \label{pulse-velocity-bianry}
c(t)\sim t^{-f(q)/5}.
\end{equation}
Thus, for backward exponentially TCs the pulse amplitude and position 
vary exponentially in $k$, but in time the variations are power laws.
\begin{figure}[h]
\centering
\rotatebox{0}{\scalebox{.30}{\includegraphics{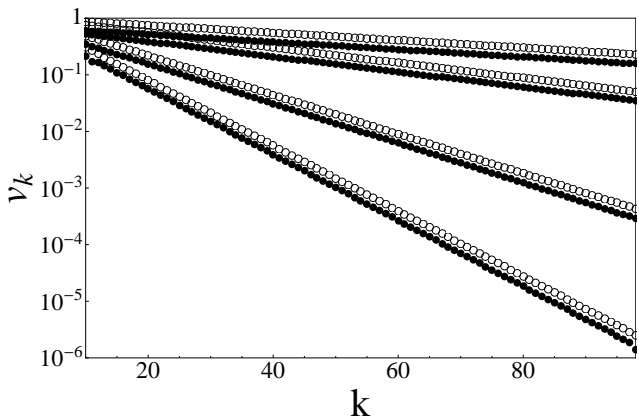}}}
\caption{The numerical data (solid circles) is compared with the binary collision approximation
(open circles) for the decay of the velocity pulse amplitude in backward exponentially TCs
for $q=0.01$, $0.02$, $0.05$, and $0.08$ from top to bottom.}
\label{figure4}
\end{figure}
\begin{table}
  \centering
\begin{tabular}{|l|l|l|}
  \hline
  q & theory & numerics\\ \hline
  0.01 &  0.01519 &  0.01522\\ \hline
  0.02 &  0.03076 &  0.03088\\ \hline
 0.03 & 0.04673 & 0.04697\\ \hline
0.04 & 0.06311 & 0.06350\\ \hline
 0.05 &  0.07990 &  0.08047\\ \hline
0.06 &  0.09711 &  0.09788\\ \hline
0.07 &  0.11477 &  0.11568\\ \hline
  0.08 &  0.13287 &  0.13378\\ \hline
0.09 &  0.15144 &  0.15275\\ \hline
\end{tabular}
 \caption{Exponential rates of decay for the velocity pulse in $k$-space
for backward exponentially TCs as obtained from the binary collision
approximation [$\ln A(q)$ in Eq.~(\ref{eq:vgeometric})]
 and the numerical integration of Eq.~(\ref{eq:motion_rescaled}).}
  \label{tab:kexp}
\end{table}

Figure~\ref{figure4} shows the exponential decay of the pulse amplitude with $k$ as predicted from
the theory, Eq.~(\ref{eq:vgeometric}), and as obtained from direct numerical integration of the
equations of motion.  We find a small difference between the amplitudes
predicted by theory and those obtained from the numerical integration,
but the predicted decay rate is again in excellent agreement with the 
actual data. We compare the decay rates for each $q$ value obtained by fitting the
numerical data in Fig.~\ref{figure4} with those calculated from the theory.
The results are shown in Table II.

\begin{figure}[h]
\centering
\rotatebox{0}{\scalebox{.30}{\includegraphics{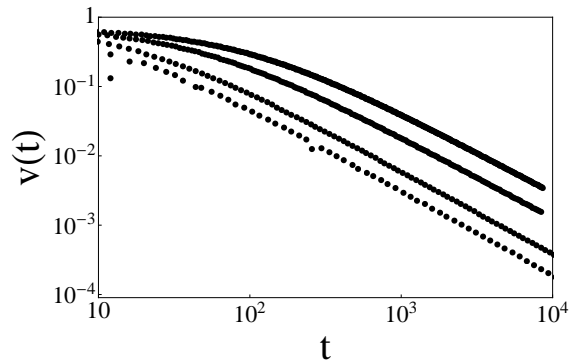}}}
\caption{Decay of the pulse amplitude with time for backward exponentially TCs
obtained from numerical integration
for $q=0.01$, $0.02$, $0.05$, and $0.08$ from top to bottom.}
\label{figure5}
\end{figure}

The power law decay of the pulse amplitude in time is shown in Fig.~\ref{figure5}. The rates
of decay are close to those
predicted by the theory, Eq.~(\ref{vel-t-binary}).
In Table~\ref{tab:tpowerlaw} we compare the two results for different $q$ values.
Note that the asymptotic expressions are valid
only in the long time limit, a requirement that involves longer times 
as the value of $q$ decreases. The small differences in the results arise mainly because
Eq.~(\ref{vel-t-binary}) is a long-time limit (and not because of the binary approximation per se).
\begin{table}
  \centering
\begin{tabular}{|l|l|l|}
\hline
  q & theory & numerics\\ \hline
  0.01 &  1.1605 &  1.1279\\ \hline
  0.02 &  1.1673 &  1.1546\\ \hline
0.03 &  1.1740 & 1.1744 \\ \hline
  0.04 &  1.1808 & 1.1820\\ \hline  
  0.05 &  1.1877 & 1.1803\\ \hline
  0.06 &  1.1945 & 1.1952\\ \hline
0.07 &  1.2015 &  1.2049\\ \hline
  0.08 &  1.2084 & 1.2145\\ \hline
0.09 &  1.2154 & 1.2256\\ \hline
\end{tabular}
 \caption{Asymptotic rates of decay for the pulse amplitude in time for backward
exponentially TCs obtained from the binary collision approximation [$f(q)$ in
Eq.~(\ref{bq-geometric})] and those obtained by fitting the straight
line portions of the data in Fig.~\ref{figure5}.}
  \label{tab:tpowerlaw}
\end{table}

\begin{figure}[h]
\centering
\rotatebox{0}{\scalebox{.30}{\includegraphics{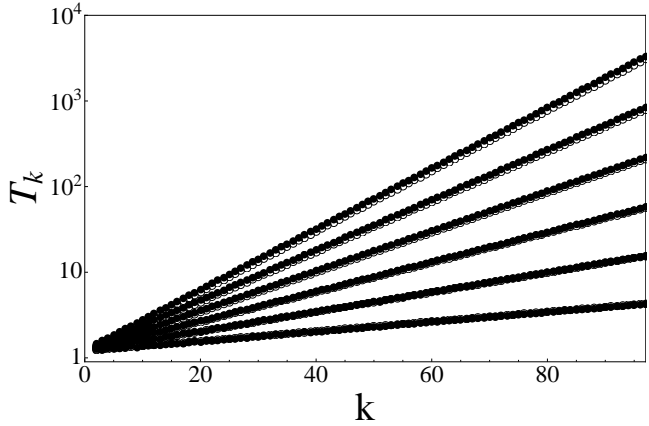}}}
\caption{The pulse residence time $T_k$ as a function of $k$
for backward exponentially TCs is shown for $q=0.01$ to $0.06$
in steps of $0.01$ (top to bottom).  Theory and numerical results are denoted
by open circles and filled circles, respectively.}
\label{figure5a}
\end{figure}

\begin{figure}[h]
\centering
\rotatebox{0}{\scalebox{.30}{\includegraphics{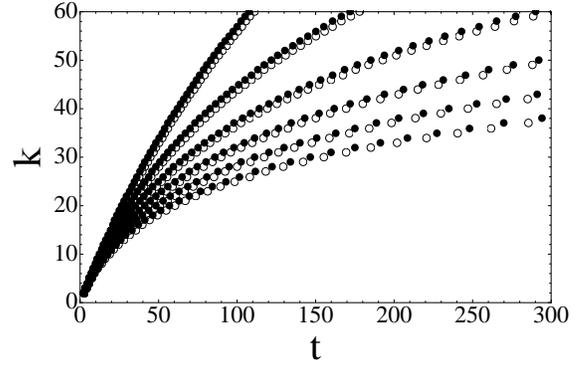}}}
\caption{The position of the pulse in units of grains versus time $t$
for backward exponentially TCs is shown for $q=0.01$ to $0.06$ in steps of $0.01$
(top to bottom).  Theory and numerical results are denoted by open circles and
filled circles, respectively.}
\label{figure6}
\end{figure}

Finally, the theoretical residence times $T_k$ and the time $t$ for the pulse to pass through the $k$th granule
are in excellent agreement with the numerical integration data, as can be seen in
Figs.~\ref{figure5a} and \ref{figure6}.

\section{Forward Tapered Chains}
\label{forward}

Next we consider chains where the granules are successively smaller starting from the granule that is the subject of the initial impulse, that is, chains with forward tapering.  We present our binary collision approximation results 
and compare them with the results obtained from the numerical 
integration of the full equations of motion (\ref{eq:motion_rescaled}). We
again consider two different tapering protocols, the linear and the exponential.

\subsection{Linearly tapered chains}

In forward linearly TCs the radii of the granules decrease linearly as $R_k=1-S(k-1)$. 
Note that since the radius of the first granule is unity, this places an $S$-dependent restriction
on the chain length $N$ (or an $N$-dependent restriction on the tapering parameter $S$),
\begin{equation}
\label{r1-forw}
N < 1+\frac{1}{S}.
\end{equation}
In our calculations we set $N=100$, and accordingly restrict the value of $S$ to be $\leq 0.01$.
The ratios of the radii and of the masses of the $k$th to ($k+1$)st granules are
\begin{eqnarray}
 \label{ratio-forw}
\frac{R_k}{R_{k+1}} &=& 1+\frac{S}{1-Sk}\\ \nonumber\\
\frac{m_k}{m_{k+1}} &=& \left(1+\frac{S}{1-Sk}\right)^3.
\end{eqnarray}
Substituting Eq.~(\ref{ratio-forw}) in Eq.~(\ref{recv}), we obtain
\begin{equation}
 \label{recv-forw}
v_k=\prod_{k^\prime=1}^{k-1} \frac{2}{1+\left(1+\frac{\displaystyle S}{\displaystyle
1-Sk^\prime}\right)^{-3}}.
\end{equation}
When $k \ll 1+1/S$ we follow the steps described in going from Eq.~(\ref{fullvk}) to
Eq.~(\ref{vel-ampl-lin-large}) to find the scaling behavior
\begin{equation}
\label{vk-forw}
v_k \sim \left(1-\frac{k}{1+1/S}\right)^{-3/2}.
\end{equation}
Since $k< 1+1/S$, this function grows as $k$ increases. This is expected since grain mass decreases
with increasing $k$.  The associated parameters $\mu_k$ and $r_k$ for forward linear TCs
are obtained by replacing $S$ with $-S$ in Eqs.~(\ref{mk-ak-linear-1}) and
(\ref{mk-ak-linear-2}).
Using Eq.~(\ref{vk-forw}) in Eq.~(\ref{compress-time}) and summing over $k$ we obtain the
scaling for the time it takes the pulse to pass through the $k$th granule,
\begin{equation}
 \label{t-forw}
t\sim C(S) \left[ 1- (1+S)^{23/10}\left(1-\frac{k}{1+1/S}\right)^{23/10}\right],
\end{equation}
where $C(S) = C_0(S) \left[ 1+{\mathcal{O}(S)}\right]$ and
\begin{equation}
C_0(S)=
\frac{10\sqrt{\pi}}{23}\left(\frac{5}{8}\right)^{2/5}\frac{\Gamma(7/5)}{\Gamma(9/10)}
\left(1+\frac{1}{S}\right).
\end{equation}

\begin{figure}[h]
\centering
\rotatebox{0}{\scalebox{.30}{\includegraphics{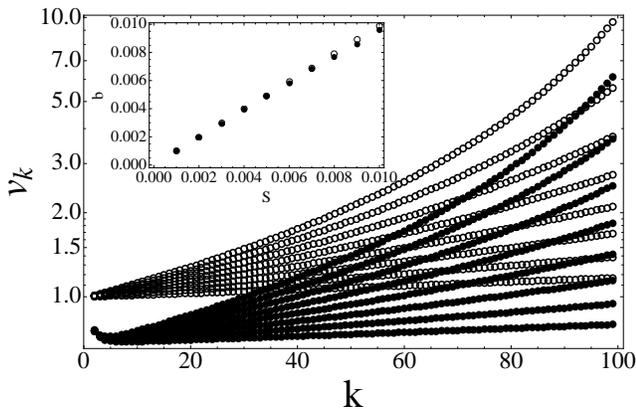}}}
\caption{The numerical integration data (filled circles) 
is compared with the binary collision approximation results (open circles) for the
velocity pulse amplitude in forward linearly TCs
for $S$ from $0.001$ to $0.008$ in steps of $0.001$ (bottom to top).
The inset shows the variation of the parameter $b$ with $S$ in the fit of $v_k$
described in the text.}
\label{figure8}
\end{figure}

Figure~\ref{figure8} shows the variation in the pulse amplitude with $k$ as obtained from
the theory (\ref{recv-forw}) and the numerical integration. 
As predicted from Eq.~(\ref{vk-forw}), we observe a linear growth in the
pulse amplitude for small $k$. 
Also, as $k \to N$ the numerical results show that the velocity behaves as
$v_k \sim a(1-bk)^{-3/2}$ and thus diverges around $k\sim 1/b$. Interestingly, we find that $1/b$ is
very close to $(1+1/S)$ and thus Eq.~(\ref{vk-forw}), which is in principle restricted to the regime $k\ll
1+1/S$, seems to work well all the way up to the divergence. 
In the inset in Fig.~\ref{figure8} we compare
the values of $b$ obtained from the theory, $b=(1+1/S)^{-1}$, and the fit to numerical data. 

\begin{figure}[h]
\centering
\rotatebox{0}{\scalebox{.30}{\includegraphics{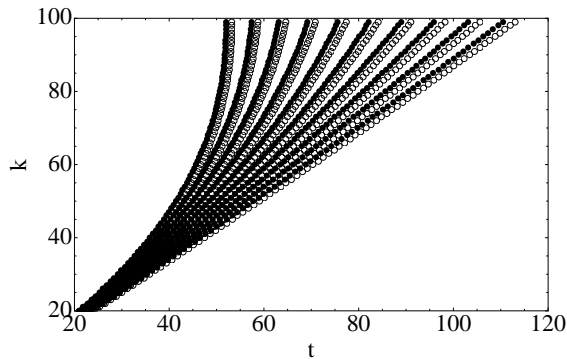}}}
\caption{Comparison of the theory (open circles) and numerical integration results
(filled circles) for the time $t$ taken by the pulse
in forward linearly TCs to reach
the $k$th granule for $S$ from $0.001$ to $0.01$ in steps of $0.001$ from
right to left.}
\label{figure9}
\end{figure}

In Fig.~\ref{figure9}, we compare the results for the time $t$ taken
by the pulse to pass through the $k$th granule.
The change in $t$ with $k$ is described extremely accurately by the binary
collision approximation result~(\ref{t-forw}) valid for $k\ll 1/S$.
The result in Eq.~(\ref{t-forw}) is thus expected to be valid for all $k$ provided that $S$ is small
($1/S\gg 1$).  Interestingly, the result in Eq.~(\ref{t-forw}) in fact works well for all
$k\leq N$ (with $N=100$) in the range $0.001 \leq S \leq 0.01$.  The time $t$ increases linearly
with $k$ for small $k$ but becomes almost $k$-independent as $k$ increases.  This is a reflection of
the fact that the granular mass and size become very small and as a result the pulse passes through
these grains very quickly.

\begin{figure}[h]
\centering
\rotatebox{0}{\scalebox{.30}{\includegraphics{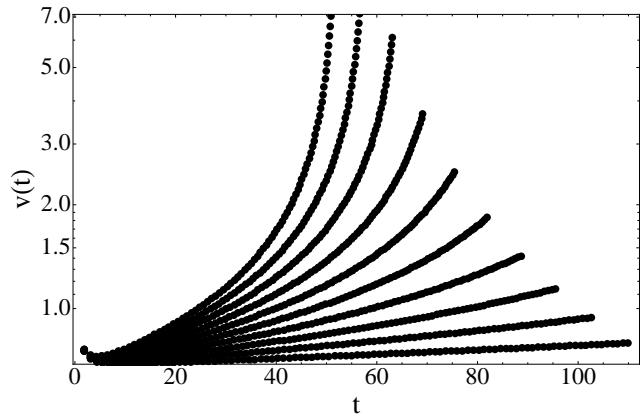}}}
\caption{Change in the pulse amplitude with time for forward linearly TCs.
as obtained from numerical integration of the equations of motion.
Results are for $S$ from $0.001$ to $0.01$ in steps
of $0.001$ from bottom to top.}
\label{figure10}
\end{figure}

\begin{figure}[h]
\centering
\rotatebox{0}{\scalebox{.30}{\includegraphics{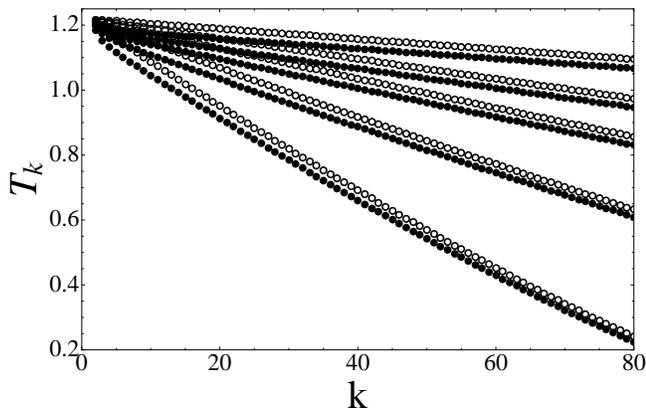}}}
\caption{Residence time of the pulse on each granule
for forward linearly TCs for $S=0.001$, $0.002$, $0.003$, $0.005$, and $0.009$
from top to bottom. Numerical integration results: filled circles; binary
collision approximation results:
open circles.}
\label{figure11}
\end{figure}

The change in the pulse amplitude with time is shown in Fig.~\ref{figure10}.
The theoretical result for the time-dependent amplitude can be obtained by combining
Eq.~(\ref{vk-forw}) with Eq.~(\ref{t-forw}). 
We find that
\begin{equation}
\label{v-t-forw}
v(t) \sim \left( 1-\frac{t}{C(S)}\right)^{-15/23}.
\end{equation}
For small $t$ the pulse amplitude thus grows linearly with $t$, and as $t\to C(s)$ it diverges with
exponent $15/23$. Since the absolute value of the pulse amplitude is not captured correctly by
the binary collision approximation, the divergence time $C(S)$ of the pulse amplitude is
also different from the
actual divergence time seen in the numerics.  However, we find that
the initially linear behavior and the divergence exponent $15/23$ are
accurately predicted by the theory.  In Fig.~\ref{figure11} we show the 
pulse residence time obtained for each granule from numerical integration
along with the corresponding values obtained from the binary collision approximation. 

\subsection{Exponentially tapered chains}

\begin{figure}[h]
\centering
\rotatebox{0}{\scalebox{.30}{\includegraphics{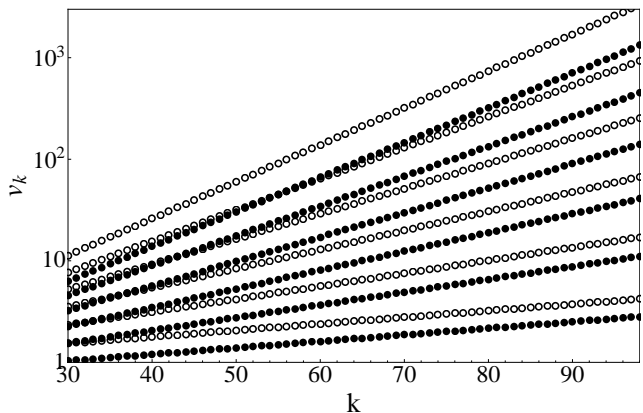}}}
\caption{Change in the pulse amplitude as a function of $k$
for backward exponentially TCs for $q$ from $0.01$ to $0.06$ in steps
of $0.01$ from bottom to top. Numerical integration results: filled circles. Binary
collision approximation results: open circles.}
\label{figure12}
\end{figure}

In this case the radii of the granules decrease as
\begin{eqnarray}
 \label{rk-geo-forw}
R_k=\frac{1}{(1+q)^{k-1}}
\end{eqnarray}
where $q>0$. The velocity of the $k$th granule can be obtained from
Eq.~(\ref{eq:vgeometric}) by replacing $q$ by $-q$. Figure~\ref{figure12} shows the
comparison of the binary collision approximation results with
the numerical data. We again observe that although the
absolute values differ by a constant amount, the rate of growth is 
correctly predicted by the theory. 
In Table~\ref{tab1:geo:forvk} we compare the rates obtained from the fit to
the data points in Fig.~\ref{figure12} with the results of the theory.
\begin{table}
  \centering
\begin{tabular}{|l|l|l|}
\hline
  q & theory & numerics\\ \hline
  0.01 & 0.01481 & 0.01471\\ \hline
  0.02 & 0.02926 &  0.02879\\ \hline
  0.03 &  0.04336 & 0.04267 \\ \hline
  0.04 & 0.05710 & 0.05565\\ \hline  
  0.05 & 0.06848 & 0.07051\\ \hline
  0.06 & 0.08359 & 0.07833\\ \hline
\end{tabular}
 \caption{Comparison of the rates for the exponential growth of the pulse amplitude
in forward exponentially TCs in granule space obtained from the binary collision
approximation with those obtained by fitting the data in Fig.~\ref{figure12}.}
  \label{tab1:geo:forvk}
\end{table}

\begin{figure}[h]
\centering
\rotatebox{0}{\scalebox{.40}{\includegraphics{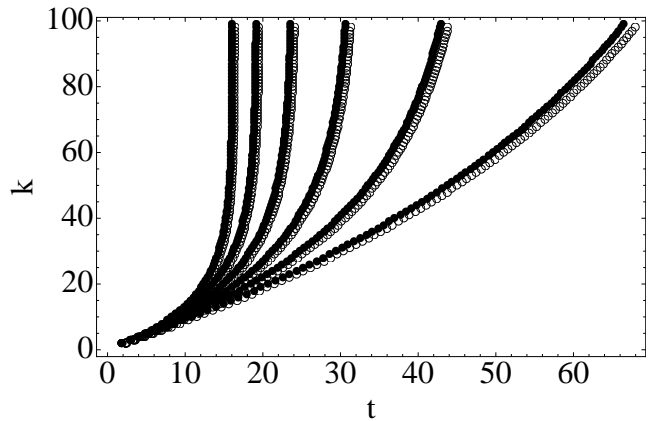}}}
\caption{Total time taken by the pulse to pass through the $k$th granule
in forward exponentially TCs for
$q$ from $0.01$ to $0.06$ in steps of $0.01$
(right to left).  Numerical integration results: filled circles. Binary
collision approximation results: open circles.}
\label{figure13}
\end{figure}

The total time taken by the pulse to reach the $k$th granule is obtained by
substituting Eqs.~(\ref{eq:vgeometric}) and (\ref{mu-binary})
in Eq.~(\ref{compress-time}), replacing  $q$ by $-q$, and summing over $k$, cf. Eq.~(\ref{totaltime}).
For long times, we obtain
\begin{eqnarray}
 \label{geo-forw--t-k}
t\sim 1-\mbox{e}^{(k-1)\eta(-q)},
\end{eqnarray}
where $\eta(-q)$ is obtained by replacing $q$ by $-q$ in Eq.~(\ref{Bq}).
Since $\eta(-q)<0$, $t$ approaches a constant value at large $k$. In Fig.~\ref{figure13}
we show both the theory and the numerical results for the time taken by the pulse to pass
through the $k$th granule.

\begin{figure}[h]
\centering
\rotatebox{0}{\scalebox{.30}{\includegraphics{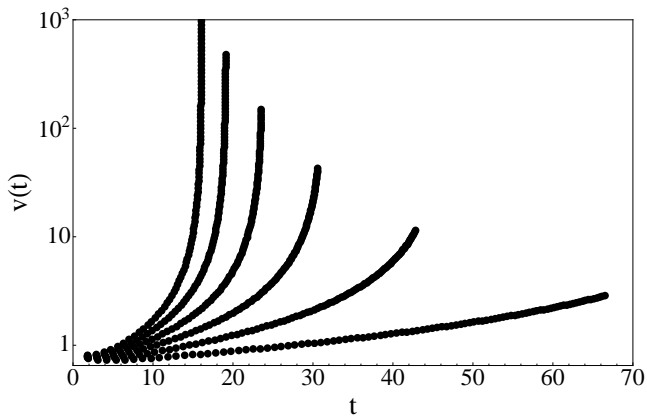}}}
\caption{Numerical integration results for the pulse amplitude as a function of time
in forward exponentially TCs for $q$ from $0.01$ to $0.06$ in steps of $0.01$ (bottom to top).}
\label{figure14}
\end{figure}

The time dependence of the pulse amplitude in the binary collision approximation is obtained as
\begin{equation}
 \label{geo-vt-forw}
v(t)\sim \left(1+\frac{t}{M(q)}\right)^{-f(-q)},
\end{equation}
where 
\begin{equation}
 \label{mq}
M(q) = \frac{2\sqrt{\pi}}{5\eta(-q)}\frac{\Gamma(7/5)}{\Gamma(9/10)}
\left(\frac{5}{4} \frac{\sqrt{1-q/2}}{1+(1+q)^3}\right)^{2/5},
\end{equation}
and $f(-q)$ is obtained by replacing $q$ with $-q$ in Eq.~(\ref{bq-geometric}). 
At short times $v(t)$ increases linearly with $t$.  Also,
since $\eta(-q)<0$, $M(q)<0$ and, as
$t\to M(q)$, the velocity diverges with the exponent $f(-q)$. These behaviors are
shown in Fig.~\ref{figure14}. Again, since the absolute value of the pulse
amplitude is not captured correctly within the binary collision approximation, the
divergence time of the pulse
amplitude $M(q)$ is also different from the actual
divergence time seen in the numerics. However, the exponent $f(-q)$ is accurately predicted
by the theory. In Table~\ref{tab:geo:vt-powerlaw}
\begin{table}
  \centering
\begin{tabular}{|l|l|l|}
\hline
  q & theory & numerics\\ \hline
  0.01 & 1.1472 & 1.1361\\ \hline
  0.02 & 1.1406 & 1.1308\\ \hline
  0.03 & 1.1341 & 1.1134 \\ \hline
  0.04 & 1.1276& 1.1237\\ \hline  
  0.05 & 1.1211 & 1.1022\\ \hline
  0.06 & 1.1147 & 1.1101\\ \hline
\end{tabular}
 \caption{Comparison of the power law exponent $f(-q)$, Eq.~(\ref{geo-vt-forw}), obtained
from the binary collision approximation with the exponent extracted from 
numerical integration results for forward exponentially TCs.}
  \label{tab:geo:vt-powerlaw}
\end{table}
we compare the exponent $f(-q)$ obtained by fitting the numerical data in Fig.~\ref{figure14} with
the results of the theory.

\vspace{2cm}
\section{Conclusions}
\label{summary}

We have introduced an analytic approach to calculate the behavior of a
propagating pulse along a
variety of tapered chains of spherical granules.  We implement a binary collision
approximation~\cite{alexandremono} that
supposes that the collision events by which a pulse propagates involve only two granules at a time.
While the approximation overestimates the pulse amplitude, it captures all other properties
remarkably well.  These include the rate of decay of the pulse amplitude with time,
the residence time of the pulse on each granule, and the pulse speed in units of granule and in
space.

We have implemented our approximation in four different tapered chains: backward linearly and
exponentially tapered chains (the granules increase in size), and forward linearly and exponentially
tapered chains (the granules decrease in size).
In backward linearly tapered chains, where granular size increases linearly along the direction
of pulse propagation, an initial impulse on 
the first granule settles into a pulse whose amplitude, speed and width change
slowly as the pulse propagates. The pulse speed and the pulse amplitude
decay in granular units and also in real time, as is reasonable since the granules
increase in size and mass along the direction of propagation.
Although the pulse remains narrow in granule number, the
pulse width in real space of course increases as the granules get larger.  All these decreases and
increases vary as power laws both in granular units $k$
and in time $t$ for large $k$ and $t$, and the decay and growth exponents are insensitive 
to the value of the tapering $S$ over the range that we have tested, $0.1<S \leq 1$.
On the other hand, for backward exponentially tapered chains, where granule size increases
geometrically, the pulse properties change exponentially in $k$ while they
exhibit power law behavior in time.  In this case the rates are strongly influenced by the
tapering parameter $q$.

In forward linearly tapered chains, where granular size decreases linearly, the pulse speeds up 
both in granule number units and in time, and in fact diverges.  The divergence exponents are
again insensitive to the tapering parameter $S$ and are
quantitatively reproduced by the theory. Similar divergences are observed in the forward
geometrically tapered chain, but now the behavior is exponentially sensitive to the tapering
parameter $q$. 

In the binary collision approximation the initial velocity for the collision between granules $k$
and $k+1$ is taken to be the velocity of granule $k$ at the end of the collision between granules
$k-1$ and $k$. This estimate in turn leads to a pulse velocity amplitude that
is higher than that obtained from the numerical integration of the full equations of motion,
but all other pulse characteristics are
quantitatively reproduced in all our tapered chains by this analytic approximation.
Our next challenge is to generalize this approach to chains with other mass
variation profiles and even to random chains. This work is in progress.

\section*{Acknowledgments}
Acknowledgment is made to the Donors of the American Chemical Society Petroleum Research Fund for
partial support of this research (K.L. and U.H.).  A.R. acknowledges support from Pronex-CNPq-FAPESQ
and CNPq.  M. E. is supported by the FNRS Belgium (charg\'e de recherches) and
by the government of Luxembourg (Bourse de formation recherches).

\end{document}